\documentclass[smallcondensed]{svjour3}     
\usepackage[english]{babel} 
\usepackage{graphicx}
\usepackage{natbib}
\usepackage[T1]{fontenc}
\usepackage[misc]{ifsym}
\begin{document}

\title{Multilayer metamaterial absorbers inspired by perfectly matched layers\thanks{This work was supported by research project UMO-2011/01/B/ST3/02281 of the Polish National Science Center.
PL-Grid infrastructure is acknowledged for providing access to computational resources.
}
}

\titlerunning{Multilayer realization of UPML}        

\author{Anna Pastuszczak \and Marcin Stolarek \and Tomasz J. Antosiewicz \and Rafa{\l} Koty{\'n}ski}

\authorrunning{A. Pastuszczak et al.} 

\institute{A. Pastuszczak  \and
           M. Stolarek \and
           R. Koty{\'n}ski {\Letter}\at
              Faculty of Physics, University of Warsaw, Pasteura 7, 02-093 Warsaw, Poland \\
              \email{rafalk@fuw.edu.pl}, Tel.: +48-225546888, Fax: +48-225546882          
           \and
           T. J. Antosiewicz \at
              Centre of New Technologies, University of Warsaw, \.Zwirki i Wigury 93, 02-089 Warsaw, Poland\\
              Department of Applied Physics and Gothenburg Physics Centre, Chalmers University of Technology, SE-41296 G\"oteborg, Sweden                 
}

\date{Received: date / Accepted: date}

\maketitle

\begin{abstract}
We derive periodic multilayer absorbers with effective uniaxial properties similar to perfectly matched layers (PML). This approximate representation of PML is based on the effective medium theory and we call it an effective medium PML (EM-PML). We compare the spatial reflection spectrum of the layered absorbers to that of a PML material and demonstrate that after neglecting gain and magnetic properties, the absorber remains functional. This opens a route to create electromagnetic absorbers for real and not only numerical applications and as an example we introduce a layered absorber for the wavelength of $8$~$\mu$m made of SiO$_2$ and NaCl.
We also show that similar cylindrical core-shell nanostructures derived from flat multilayers also exhibit very good absorptive and reflective properties despite the different geometry. 
\end{abstract}

\section{Introduction}
\label{intro}
Perfectly matched layer (PML)~\citep{PML_Berenger_2007} absorbers  are now widely used to terminate electromagnetic simulations with an open domain. PMLs suppress reflection and ensure absorption of incident electromagnetic radiation at any angle and any polarization.  A variety of PML formulations exist, starting from the early split-field PML~\citep{JCP_114_185_berenger}, and the coordinate stretching approach~\citep{IEEE_MGWL_7_599_chew} up to the convolutional PML (CPML)~\citep{MOTL_27_334_roden}, and the near PML (NPML)\citep{MWL_13_128_cummer}. In this paper we refer to the Maxwellian formulation of PML, represented by an artificial material with uniaxial permittivity and permeability tensors, usually termed as uniaxial PML (UPML)~\citep{IEEE_TAP_43_1460_sacks, IEEE_TAP_44_1630_gedney}. A PML can be used with both time-domain and frequency domain methods, as well as with finite difference or finite element discretization schemes. It can assume various dispersion models, see e.g. the time-derivative 
Lorentz material that is capable of absorbing oblique, pulsed electromagnetic radiation having narrow and broad waists~\citep{IEEE_TAP_45_656_ziolkowski}. A PML can not be applied in some rare cases and for instance it fails to absorb a backward propagating wave for which an adiabatic absorber should be used instead~\citep{OE_16_11376_Zhang,PRE_79_065601_2009_loh}.

Electromagnetic absorbers have a much longer history than any kind of numerical modeling. Their possible applications range from modification of radar echo, through applications related to electromagnetic compatibility, up to photovoltaics. Early real-world absorbers were based on resistive sheets separated from a ground plate by quarter wave distances. With several sheets and multiple resonances it was possible to achieve broadband operation. The idea evolved into the theory of frequency selective surfaces~\citep{Munk_FSS}. Furthermore, it is possible to obtain a tailored impedance at a surface transition region using homogenized periodic one-dimensional or two-dimensional corrugated surfaces~\citep{PIER_55_1_Kristensson}. A static periodic magnetization obtained with ferromagnetic or ferrimagnetic materials is another route to obtaining broadband absorbers~\citep{PIER_83_199_Ramprecht}. A recent overview paper~\citep{AMO_24_98_watts} can serve as a tutorial on absorbers with the focus on novel  
metamaterial absorbers based on split-ring and electric-ring resonators. 

In this paper we introduce the effective medium PML absorbers (EM-PML), which are metamaterial absorbers with a layered structure that exhibit effective permittivity and permeability tensors similar to a PML material. We calculate the reflection coefficient achieved with these layered absorbers. We look towards their possible physical realizations.

\begin{figure}
\centering
\includegraphics{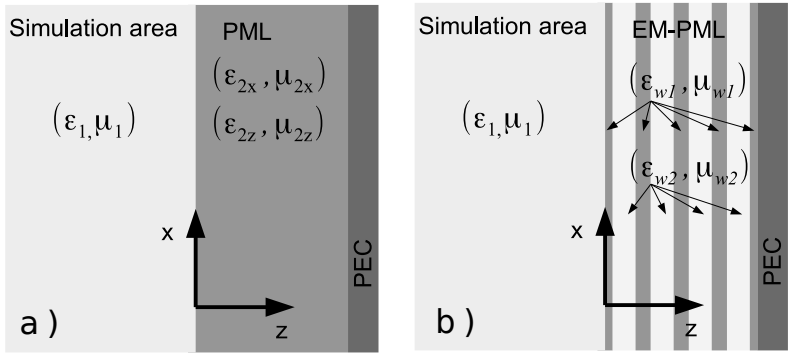} 
\caption{ a)~A uniaxial PML characterized by anisotropic permittivity and permeability tensors is attached to a simulation area ($\varepsilon_{1}, \mu_{1}$). b) An effective-medium PML, which has the same uniaxial properties, can be composed of two isotropic materials ($\varepsilon_{wi}, \mu_{wi}$, where $i=1,2$) arranged in a multilayered fashion. In both cases a perfect electric conductor (PEC) terminates the PML 
\label{fig.schem}}
\end{figure}

\section{Approximate representation of UPML}
\label{sec:1}
A schematic of a uniaxial perfectly matched layer attached to a simulation area is shown in Fig.~\ref{fig.schem}. In the outer area of the simulation domain (which neighbors the PML), the permittivity and permeability are equal to $\varepsilon_1$, and $\mu_1$. The UPML is defined as a material whose permittivity $\varepsilon_2$ and permeability $\mu_2$  take the following tensor forms,
\begin{equation}
\varepsilon_2 =\left[ \begin{array}{ccc}
\varepsilon_{2x}&0&0\\ 0&\varepsilon_{2x}&0\\0&0&\varepsilon_{2z}
\end{array}\right], \quad
\mu_2 = \left[\begin{array}{ccc}
\mu_{2x}&0&0\\ 0&\mu_{2x}&0\\0&0&\mu_{2z}
\end{array}\right] \label{eq.upml},
\end{equation}
where
\begin{equation}
 \varepsilon_{2x} = s\cdot \varepsilon_{1}, \quad \mu_{2x} = s\cdot\mu_{1}  , \quad
\varepsilon_{2z} = s^{-1} \cdot\varepsilon_{1} ,  \label{eq.tm}
\end{equation}
and
\begin{equation}
\mu_{2x} = s\cdot\mu_{1}, \quad \varepsilon_{2x} = s\cdot\varepsilon_{1}, \quad
 \mu_{2z} = s^{-1} \cdot\mu_1.\label{eq.te}
\end{equation}
The parameter $s$ is a non-zero freely chosen complex number, whose imaginary part determines the strength of absorption within the UPML. The conditions stated in Eq.~(\ref{eq.tm}) alone are sufficient to remove reflections for the TM polarization, for any  plane-wave propagating in-plane, independently of its angle of incidence and frequency. Equation~(\ref{eq.te}) assures the same for the TE polarization and when both conditions are satisfied, the UPML is reflection-free for any polarization and any angle of incidence in a three-dimensional case. Subsequently, $s$ may be varied with the coordinate $z$ to give a continuously graded permittivity and permeability, which works better with most discretization schemes. Then, the same parameter $s$ may also be  linked to a coordinate mapping from real to complex coordinates.

Our approximate representation of a UPML consists of a one-dimensional stack of uniform layers. According to the effective medium theory (EMT) a stack consisting of thin layers may be homogenized and replaced by a uniform uniaxial medium with effective permittivity and permeability tensors.  For the TM polarization, the effective permittivity and permeabilty tensors of a stack consisting of two materials with permittivity and permeability pairs $(\varepsilon_{w1},\mu_{w1})$, and $(\varepsilon_{w2},\mu_{w2})$  match that of a UMPL, expressed by Eq.~(\ref{eq.upml}), when
\begin{equation}
f\cdot\varepsilon_{w1} + (1-f)\cdot\varepsilon_{w2} = s\cdot\varepsilon_1,\label{eq.c1} 
\end{equation}
\begin{equation}
[f\cdot\varepsilon_{w1}^{-1} + (1-f)\cdot\varepsilon_{w2}^{-1}]^{-1} = s^{-1} \cdot\varepsilon_1,\label{eq.c2}
\end{equation}
\begin{equation}
f\cdot\mu_{w1} + (1-f)\cdot\mu_{w2} = s\cdot\mu_1,\label{eq.c3}
\end{equation}
where $f$ is the filling factor, i.e. the volume fraction of material $w1$ in the stack. A similar condition applies for the TE polarization.
Solving equations (\ref{eq.c1}) and (\ref{eq.c2}) for $\varepsilon_{w1}$, and $\varepsilon_{w2}$ yields,
\begin{equation}
 \varepsilon_{w1} = \rho\cdot\frac{\varepsilon_{1}\cdot s }{f \cdot\rho + (1-f)},\quad
 \varepsilon_{w2} = \frac{\varepsilon_{1}\cdot s }{f \cdot\rho + (1-f)},\label{eq.w1w2}
\end{equation}
where
\begin{equation}
 \rho= 1 + \frac{s^2-1 \pm \sqrt{(s^2-1)(s^2 - (2f-1)^2)}}{2f(1-f)}.\label{eq.ro}
\end{equation}

\begin{figure}
\includegraphics[width=12cm]{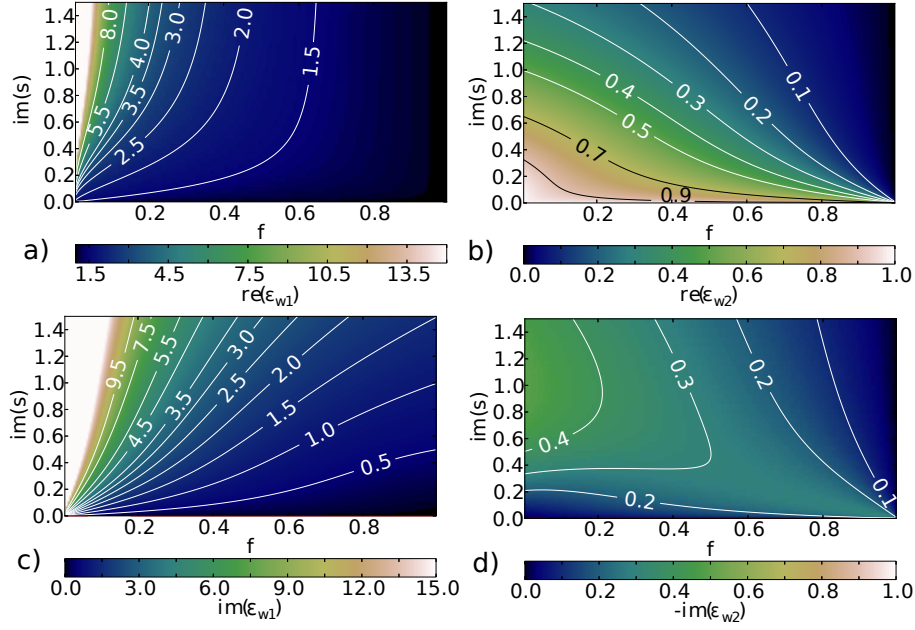}
\caption{Permittivities of the two constituent isotropic media ($\varepsilon_{w1}$ and $\varepsilon_{w2}$ in left and right columns, respectively) that form the uniaxial EM-PML as functions of the filling factor $f$ and the imaginary part of $s$ (for $re(s)=1$). Top row: real parts (a) $re(\varepsilon_{w1})$ and (b) $re(\varepsilon_{w2})$. Bottom row: imaginary parts (c) $im(\varepsilon_{w1})$ and (d) $-im(\varepsilon_{w2})$. Negative imaginary part of permittivity refers to materials with optical gain. Qualitatively, one of the materials is a high-loss material with a large refractive index (greater than one), while the other is a low-gain medium with a refractive index between $0$ and $1$\label{fig.params}}
\end{figure}

In Fig.~\ref{fig.params} we illustrate the permittivities calculated from eq.~(\ref{eq.w1w2}) as a function of the fill factor $f$ and $s$, for $s=1+\alpha i$. We use the branch of Eq.~(\ref{eq.ro}) with $|\rho|>1$.
Expressions similar to Eqs.~(\ref{eq.c1}) and~(\ref{eq.c2}) may be written and solved for $\mu_{w1}$, and $\mu_{w2}$, and when $\varepsilon_1=\mu_1$ then $\varepsilon_{w1}=\mu_{w1}$ and $\varepsilon_{w2}=\mu_{w2}$.
\begin{figure}
\centering
\includegraphics[width=12cm]{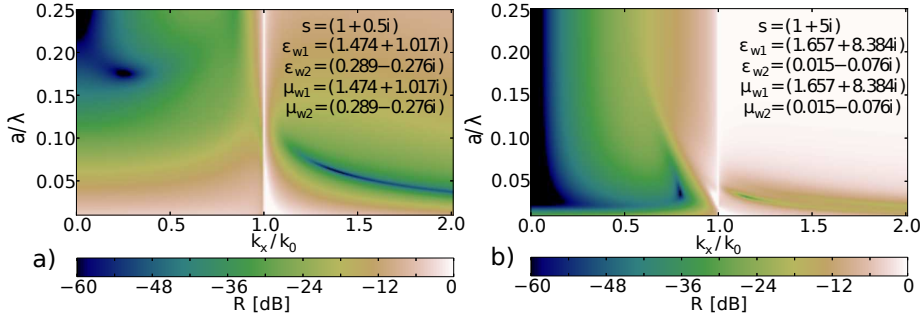}
\caption{Dependence of spatial reflection spectrum $R(k_x/k_0)$ on $a/\lambda$ for a multilayer consisting of $N=5$ periods obtained with $f=0.6$ and a)~$s=1+0.5i$, b)~$s=1+5i$. The result is polarization invariant\label{fig.perfects_pml}}
\end{figure}

\begin{figure}
\centering
\includegraphics[width=7.5cm]{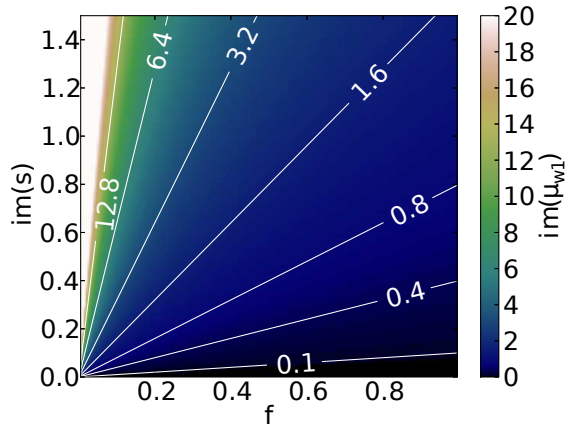}
\caption{Imaginary part of permeability of one of the materials of the EM-PML $\mu_{w1}$ as a function of the filling factor $f$ and the imaginary part of $s$ (for $re(s)=1$). Here, we have assumed that the premeability of the second material is unity $\mu_{w2}=1$, as can be done for TM polarized light, and $re(\mu_{w1})=1$ \label{fig.params_mu}}
\end{figure}

An absorber designed for the TM polarization may have one of the permeabilities freely assigned, e.g. $\mu_{w2}=1$,  while the other is then given by Eq.~(\ref{eq.c3}). If additionally $s=1+\alpha i$ then $re(\mu_{w2})=1$ and the imaginary part of $\mu_{w2}$ for this case is still positive, as shown in Fig.~\ref{fig.params_mu}. The imaginary part of $\mu_{w2}$ is small when $f$ is large and $\alpha$ is small. For either large $f$ or small $\alpha$  the negative conductivity  of the second material is also negligible (see Fig.~\ref{fig.params}d). 

The performance of a multilayer absorber obtained for $f=0.6$ and $s=1+0.5i$, and $s=1+5i$ is illustrated in Fig.~\ref{fig.perfects_pml}. Either of the two values of $s$ enables to construct an efficient broadband absorber which is at the same time subwavelength in size. Layers have both magnetic and electric properties, including gain, and a complex permeability. In the limit of $a/\lambda\rightarrow 0$ the multilayer approaches the properties of a true UPML (but at the same time, its thickness approaches $N\cdot a\rightarrow 0$). 
When $s=1+5i$, an absorber consisting of $N=5$ periods, with a total thickness of $L=5 a\approx\lambda/20$ reflects $-30dB$ for a broad range of incidence angles, and the reflection decreases rapidly with total thickness $L/\lambda$.
However, the evanescent waves are amplified in this situation. If the absorbing power is smaller, e.g. $s=1+0.5i$, the thickness $L/\lambda$ has to be larger, but the reflection is less sensitive to the magnetic permeability and gain.

\begin{figure}
\includegraphics[width=12cm]{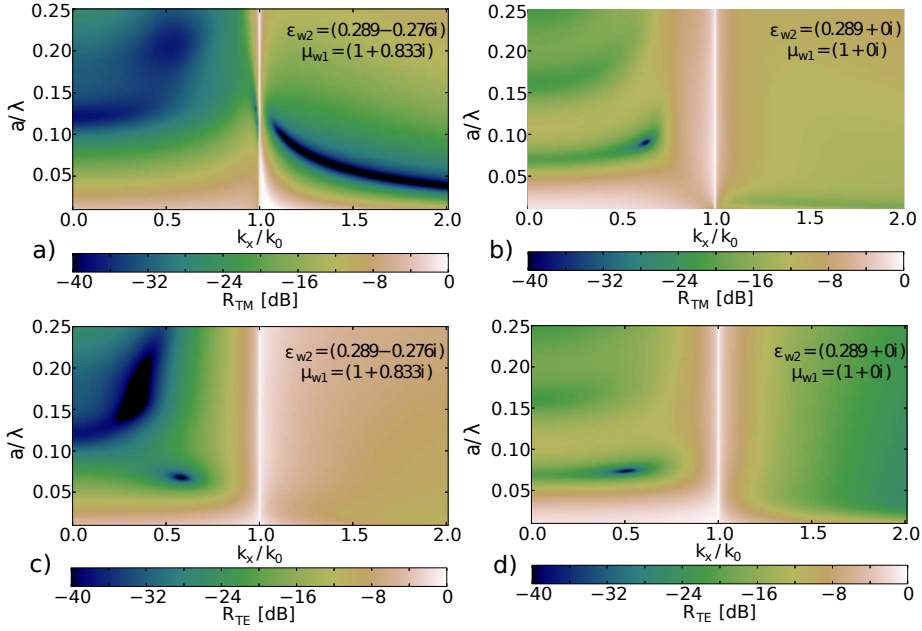}
\caption{Dependence of spatial reflection spectrum $R(k_x/k_0)$ on $a/\lambda$ for a multilayer consisting of $N=5$ periods obtained for $f=0.6$, $s=1+0.5i$ with the assumption that $\mu_{w2}=1$ (a,c), and with the assumptions that $\mu_{w1}=\mu_{w2}=1$ and $im(\varepsilon_{w1})\geq0$, $im(\varepsilon_{w2})\geq 0$ (b,d). Reflection is calculated for the TM (a,b) and TE (c,d) polarizations. The permittivity $\varepsilon_{w1}$ equals $1.474+1.017i$ \label{fig.approx_pml}}
\end{figure}

Let us asses how the reflection spectrum is changed after the magnetic permeabilities and gain have been neglected. The result is depicted in Fig.~\ref{fig.approx_pml} for $s=1+0.5i$ and $f=0.6$. The absorber consists of layers made of a lossy dielectric and of another material with permittivity lower than one.  A probable route to implement it physically is to use a metamaterial, e.g. a fishnet structure, to make use of electromagnetic mixing rules, or to use some material near its resonance frequency.  Now the reflection coefficient depends on polarization. Before neglecting gain and $im(\mu_{w1})$, the reflection has been smaller for the TM than for the TE polarization, both for the propagating and for evanescent waves. Reflection remained large only at grazing incident angles, like for ordinary UPML. However, the variant of the stack with no gain and with no magnetic properties performs better for the TE polarization (See Fig.~\ref{fig.approx_pml}bd).

Finally, the multilayer considered here has an elliptical effective dispersion relation, while similar absorbers made of hyperbolic metamaterials have been also recently proposed~\citep{PRB_86_205130_guclu}, although with no relation to PML.

\begin{figure}
\centering
\includegraphics[width=11cm]{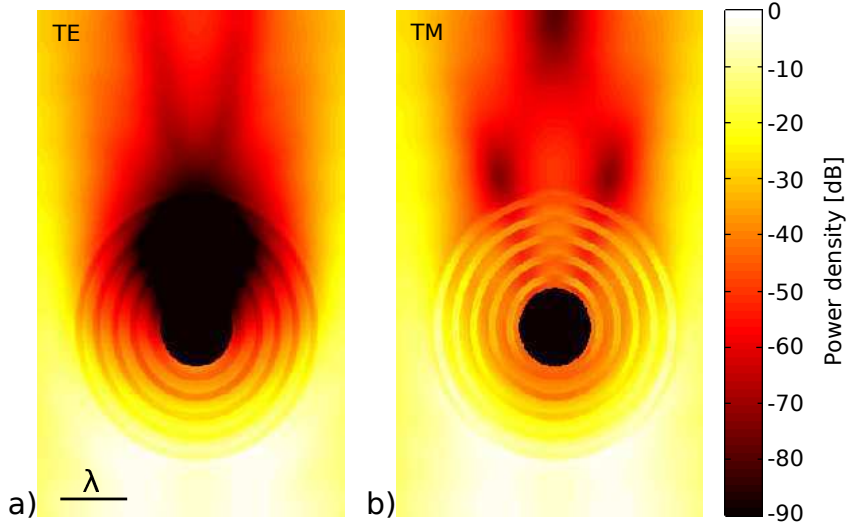}
\caption{FDTD simulation of the operation of a nonmagnetic layered absorber deposited on a cylindrical metallic object for the a)~TE-polarization, and b)~TM-polarization. The scalebar of $1\lambda$ is included in the figure. The composition of the absorber is the same as in Figs.~\ref{fig.approx_pml}cd, with $a/\lambda=0.18$ \label{fig.fdtd}}
\end{figure} 

\section{Layered slab and core-shell metamaterial absorbers}
\label{sec:2}

Based on the theoretical considerations and material parameters used to calculate the spatial reflection spectrum in Fig.~\ref{fig.approx_pml}, a simple rule of thumb for the range of required permittivities can be drawn up. This simple rule requires one permittivity to have its real part between $0$ and $1$, while the other premittivity would have its real part larger than one. The calculations show, that losses should be provided by the second material (with $\mathrm{Re}(\epsilon)>1$), while in the first material we merely neglect gain. Materials in general have $\mathrm{Re}(\epsilon)>1$, with the exception of localized transitions and broader frequency ranges in metals up to the plasma frequency.

We will now demonstrate the operation of the proposed layered absorber consisting of nonmagnetic layers with no gain.
In Fig.~\ref{fig.fdtd} we present the results of a finite difference time domain (FDTD) simulation of a layered absorber rolled into a cylindrical core-shell multilayer. The images show the time-averaged energy density distribution obtained for the TE polarization~(in Fig.~\ref{fig.fdtd}a) and TM polarization~(in Fig.~\ref{fig.fdtd}b). Light is incident from the bottom side. The composition of the metamaterial is the same as in Figs.~\ref{fig.approx_pml}cd, but the multilayer is deposited over a cylindrical  metallic (perfectly conducting) material. The absorber consists of $N=5$ periods of non-magnetic concentric layers, with the radial pitch equal to $a=0.18\lambda$, the radius of the internal PEC material is equal to $r_{int}=2 a$, and the external radius is equal to $r_{ext}=7a$. The permittivities of the layers are equal to $\varepsilon_{w1}=1.474 + 0.017i$, and $\varepsilon_{w2}=0.289$, and the filling factor equals $f=0.6$. Reflections, understood as the  part of the energy backscattered to the bottom 
side of the simulation area are as small as $R_{TE}=0.05\%$, and $R_{TM}=0.3\%$.
Thanks to the cylindrical geometry, it is possible to see the operation of the absorber at all possible angles of incidence at the same time. Notably the absorber performs well both for the TE and TM polarization, and in a spherical 3D core-shell geometry where there is no decoupling into these two polarizations we expect similar operation.
\begin{figure}
\centering
\includegraphics[width=8cm]{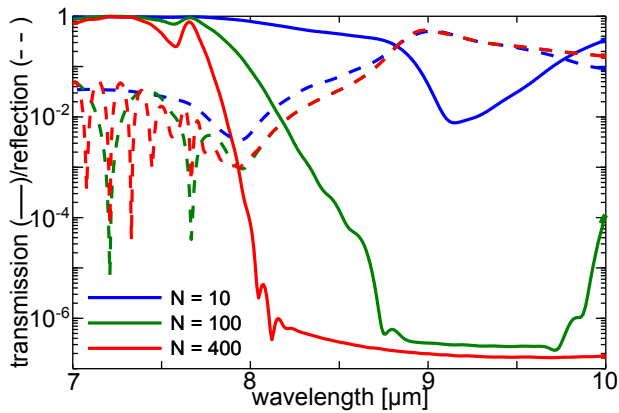}
\caption{Transmission (continuous line) and reflection (dashed) for a SiO$_{2}$/NaCl multilayer designed to absorb at 8 $\mu$m at which the refractive indices are $n_{\mathrm{SiO}_{2}}= 0.41+0.32i$, $n_{\mathrm{NaCl}}=1.51$, SiO$_2$, fill factor $f=0.56$, $a=200$ nm. The number of periods is $N=10$, 100, 400\label{fig.fdtd2}}
\end{figure}

Finally, we demonstrate the operation of a layered absorber consisting or real materials.  
Here, we make use of materials with localized electronic transitions. In the mid-infrared SiO$_2$ is such a material, which features a strong transition at approximately 9~$\mu$m and in a range between $7.2$ and $8$~$\mu$m its real part of permittivity is between $1$ and $0$. The complementary material of choice is NaCl, which in this range has $\mathrm{Re}(\epsilon)\approx1.5$, however, it also is weakly dispersive in this range and due to Kramers-Kroning relations between the real and imaginary parts of permittivity it has a very small imaginary part. Thus, SiO$_2$ needs to provide dissipation to extinguish the incident beam. Deposition of alternating SiO$_2$ and NaCl or LiF layers of required thickness may be accomplished using such techniques as chemical vapor deposition or thermal evaporation~\citep{JNCS_245_141_Fornarini,JMR_22_2012_Kim}, although with the required total thickness of the layers apporaching a few wavelengths, a fast method is preferable, at least for absorbers intended for infrared.

Fabrication of multilayered shell structures is more challenging, although synthesis of core/shell nanoparticles incorporating SiO$_2$ or Al$_2$O$_3$ and noble metals is commonplace nowadays\citep{Plasm_2014_Liu,AnChem_86_3013_Wang,JPCC_115_13660_Mai}. In the considered case here, it would be required to synthesize a layered spherical particle consisting of only dielectric materials. As mentioned previously, SiO$_2$ and Al$_2$O$_3$ have been used before, although only in relatively simple synthesis of core-shell structres; magnesium fluoride has also been synthesised in the form of nanoparticles~\citep{Nanomed_8_702_Lellouche}. However, undoubtedly some effort would be required to develop a procedure for the synthesis of layered nanoparticles.
In Fig.~\ref{fig.fdtd2} we present the results obtained for a plane wave incident normally onto a SiO$_{2}$/NaCl multilayer designed for the 7-10 $\mu$m wavelength range. The refractive index of the materials is taken from literature~\citep{palik_vol1} and the structure is optimized for $\lambda=8$~$\mu$m. The pitch is equal to $a=200$~nm, and the filling factor is $f=0.56$. Due to small losses in glass, the absorption distance is relatively long and reflection smaller than $1\%$ is observed already for 10 layers. In order to achieve both transmission and reflection smaller than $0.1\%$ $N=400$ is needed. The thickness can be reduced using  materials with larger values of the imaginary part of permittivity.

In general, a desired refractive index of one of the layers (in the range $0<\Re(n)<1$) can be manufactured using the electromagnetic mixing rules. Recently, absorbers made of hyperbolic metamaterials have been proposed~\citep{PRB_86_205130_guclu} and similar to that material, our multilayer has an elliptical dispersion with a large eccentricity.

\section{Conclusions}
\label{sec:3}
We have introduced an approximate representation of the uniaxial perfectly matched layer reflection-free absorber. The representation consists of a one-dimensional stack of uniform and isotropic metamaterial layers.  A further simplification to non-magnetic materials with no gain can be assumed for some combinations of filling fraction and absorbing power. We have also shown that similar cylindrical core-shell nanostructures derived from flat multilayers also exhibit very good absorptive and reflective properties.  A probable route to implement the absorber experimentally is by using a lossy dielectric as one material and for the other to take a metamaterial, or to make use of electromagnetic mixing rules, or to use some material near its resonance frequency. As an example we have demonstrated a layered absorber for the wavelength of $8$~$\mu$m made of SiO$_2$ and NaCl.


\end{document}